\documentclass[sigconf,screen]{acmart}

\def\BibTeX{{\rm B\kern-.05em{\sc i\kern-.025em b}\kern-.08em
    T\kern-.1667em\lower.7ex\hbox{E}\kern-.125emX}}

\AtBeginDocument{%
  \providecommand\BibTeX{{%
    \normalfont B\kern-0.5em{\scshape i\kern-0.25em b}\kern-0.8em\TeX}}}

\setcopyright{acmlicensed}
\acmDOI{10.1145/3650212.3652142}
\acmYear{2024}
\copyrightyear{2024}
\acmISBN{979-8-4007-0612-7/24/09}
\acmConference[ISSTA '24]{Proceedings of the 33rd ACM SIGSOFT International Symposium on Software Testing and Analysis}{September 16--20, 2024}{Vienna, Austria}
\acmBooktitle{Proceedings of the 33rd ACM SIGSOFT International Symposium on Software Testing and Analysis (ISSTA '24), September 16--20, 2024, Vienna, Austria}
\acmSubmissionID{issta24main-p1182-p}
\received{16-DEC-2023}
\received[accepted]{2024-03-02}

\usepackage{libertine}  
\usepackage{graphicx}
\usepackage{textcomp}
\usepackage{xcolor}
\usepackage{parallel}

\usepackage{soul}
\usepackage{subcaption}
\usepackage{url}
\usepackage{xspace}
\usepackage{enumitem}
\usepackage{multirow}
\usepackage{listings}
\usepackage{pifont}
\usepackage{booktabs}
\usepackage{array}
\usepackage[utf8]{inputenc}
\usepackage[T1]{fontenc}
\usepackage{microtype}
\usepackage{xcolor}
\newcommand{\tool}{CoEdPilot\xspace}
\newcommand{\linelocator}{\textit{Edit-propagating Line Locator}\xspace}
\newcommand{\filelocator}{\textit{Edit-propagating File Locator}\xspace}
\newcommand{\locator}{\textit{Subsequent Edit Analysis}\xspace}
\newcommand{\generator}{\textit{Edit-content Generator}\xspace}
\newcommand{\depanalyzer}{\textit{Edit-dependency Analyzer}\xspace}

\usepackage{hyperref} 
\usepackage{subfiles}

\begin{document}
\title{CoEdPilot: Recommending Code Edits with Learned Prior Edit Relevance, Project-wise Awareness, and Interactive Nature}

\author{Chenyan Liu}
\orcid{0009-0005-0554-4028}
\affiliation{%
  \institution{Shanghai Jiao Tong University}
  \city{Shanghai}
  \country{China}
}
\affiliation{%
  \institution{National University of Singapore}
  \city{Singapore}
  \country{Singapore}
}
\email{chenyan@u.nus.edu}
\authornotemark[1]

\author{Yufan Cai}
\orcid{0009-0008-7579-0824}
\affiliation{%
  \institution{Shanghai Jiao Tong University}
  \city{Shanghai}
  \country{China}
}
\affiliation{%
  \institution{National University of Singapore}
  \city{Singapore}
  \country{Singapore}
}
\email{cai_yufan@u.nus.edu}
\authornote{Both authors contributed equally to the paper}

\author{Yun Lin}
\orcid{0000-0001-8255-0118}
\affiliation{%
  \institution{Shanghai Jiao Tong University}
  \city{Shanghai}
  \country{China}
}
\email{lin_yun@sjtu.edu.cn}
\authornote{Corresponding author}

\author{Yuhuan Huang}
\orcid{0009-0001-9809-6658}
\affiliation{%
  \institution{Shanghai Jiao Tong University}
  \city{Shanghai}
  \country{China}
}
\email{hyh0u0@sjtu.edu.cn}

\author{Yunrui Pei}
\orcid{0009-0000-5624-0853}
\affiliation{%
  \institution{Shanghai Jiao Tong University}
  \city{Shanghai}
  \country{China}
}
\email{yunruipei@sjtu.edu.cn}

\author{Bo Jiang}
\orcid{0009-0000-1080-3278}
\affiliation{%
  \institution{Bytedance Network Technology}
  \city{Beijing}
  \country{China}
}
\email{jiangbo.jacob@bytedance.com}

\author{Ping Yang}
\orcid{0009-0003-9862-6983}
\affiliation{%
  \institution{Bytedance Network Technology}
  \city{Beijing}
  \country{China}
}
\email{yangping.cser@bytedance.com}

\author{Jin Song Dong}
\orcid{0000-0002-6512-8326}
\affiliation{%
  \institution{National University of Singapore}
  \city{Singapore}
  \country{Singapore}
}
\email{dcsdjs@nus.edu.sg}

\author{Hong Mei}
\orcid{0000-0003-2380-3976}
\affiliation{%
  \institution{Shanghai Jiao Tong University}
  \city{Shanghai}
  \country{China}
}
\email{meih@pku.edu.cn}


\begin{abstract}
  Recent years have seen the development of LLM-based code generation.
  Compared to generating code in a software project,
  incremental code edits
  are empirically observed to be more frequent.
  The emerging code editing approaches usually formulate the problem as generating an edit based on \textit{known} relevant
  prior edits and context. 
  However, practical code edits can be more complicated.
  First,
  an editing session can include multiple (ir)relevant edits to the code under edit.
  Second, the inference of the subsequent edits is non-trivial as the scope of its ripple effect can be the whole project.

  In this work, we propose \tool, an LLM-driven solution to recommend code edits by
  discriminating the relevant edits, exploring their interactive natures,
  and estimating its ripple effect in the project.
  Specifically, 
  \tool orchestrates multiple neural transformers to identify \textit{what} and \textit{how} to edit in the project regarding both edit location and edit content.
  When a user accomplishes an edit with an optional editing description,
  an \locator first reports the most relevant files in the project with what types of edits (e.g., \textit{keep}, \textit{insert}, and \textit{replace}) can happen for each line of their code.
  Next, an \generator generates concrete edit options for the lines of code,
  regarding its relevant prior changes reported by an \depanalyzer.
  Last, both the \locator and the \generator capture relevant prior edits as feedback to readjust their recommendations.
  We train our models by collecting over 180K commits from 471 open-source projects in 5 programming languages.
  Our extensive experiments show that
  (1) \tool can well predict the edits (i.e., predicting edit location with accuracy of 70.8\%-85.3\%, and the edit content with exact match rate of 41.8\% and BLEU4 score of 60.7);
  (2) \tool can well boost existing edit generators such as GRACE and CCT5 on exact match rate by 8.57\% points and BLEU4 score by 18.08.
  Last, our user study on 18 participants with 3 editing tasks
  (1) shows that
  \tool can be effective in assisting users to edit code in comparison with Copilot, and
  (2) sheds light on the future improvement of the tool design.
  The video demonstration of our tool is available at \url{https://sites.google.com/view/coedpilot/home}.
\end{abstract}

\ccsdesc[500]{Software and its engineering~Automatic programming}
\ccsdesc[500]{Software and its engineering~Software evolution}

\keywords{code edit generation, edit location, interaction, language model}


\maketitle

\section{Introduction}

Recent years have seen the success of the application of LM (Language Model) in code generation tasks.
LM-based approaches, such as CodeBERT \cite{feng2020codebert}, GraphCodeBERT \cite{guo2020graphcodebert}, CodeT5 \cite{wang2021codet5}, Copilot \cite{copilot}, and ChatGPT \cite{openai},
dominate the code generation solutions
by translating users' description and surrounding code context to new code.
Nevertheless, compared to generating new code,
empirical observation shows that the activities of editing existing code happen more frequently \cite{latoza2010developers, kitchin2014code, mozannar2022reading}.
Empirical study on the commits in a large number of open-source projects
shows that editing behaviors take about 70\% in the commit history \cite{nguyen2013study}.

\begin{figure}[t]
  \centering
  \includegraphics[scale=0.3]{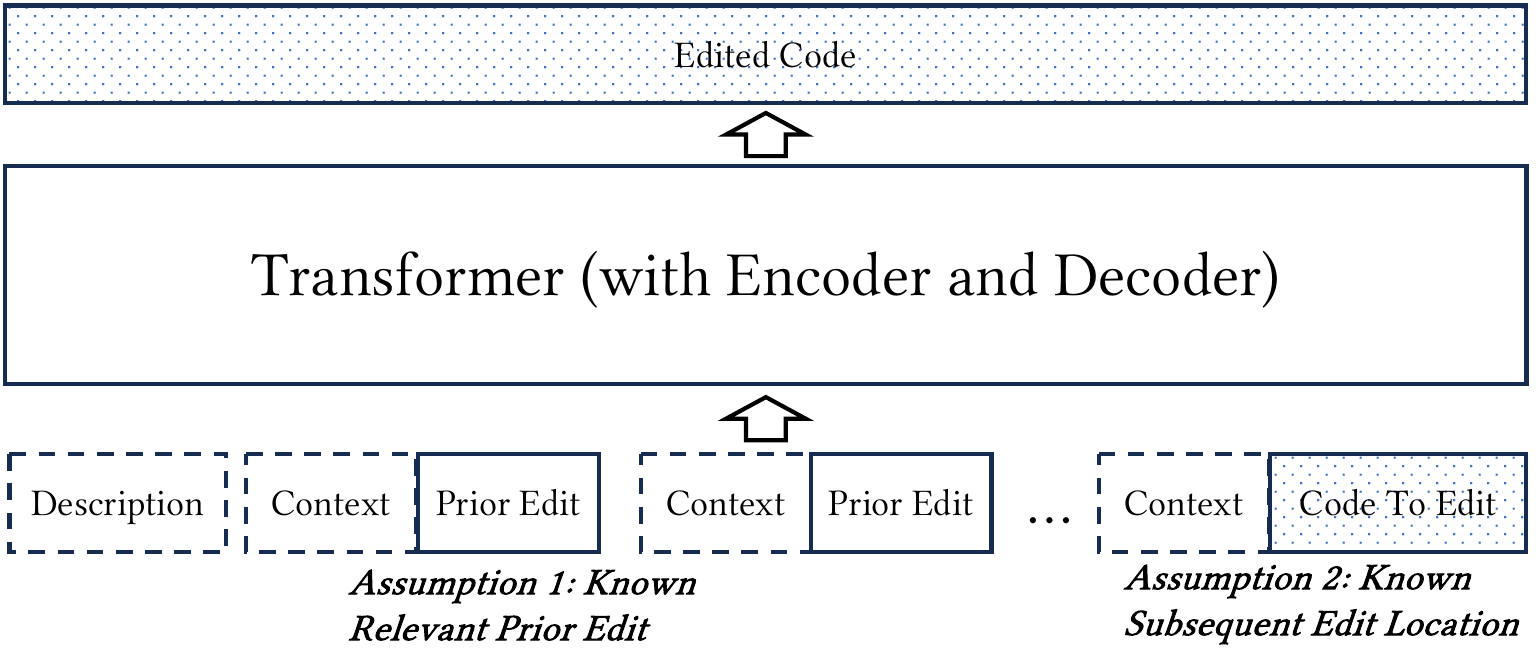}
  \caption{State-of-the-art Code Editing Framework \cite{MODIT} \cite{GRACE} \cite{CodeEditor}.
    The dotted rectangles represent the code before and after the recommended edits.}
  \label{fig:previous-work}
  \vspace{-10pt}
\end{figure}

Many transformer-based approaches are proposed to generalize
the code generation solutions to code editing solutions,
such as GRACE \cite{GRACE}, CCT5 \cite{lin2023cct5}, CoditT5 \cite{ZhangETAL22CoditT5}, and MODIT \cite{MODIT}.
While those approaches are different in representing the edits in deep learning models,
they formulate the edit generation problem as
a translation problem by
(1) taking the input as
\textit{known} relevant prior edits (and their context) and
code region (and their context) where the change is \textit{known} to happen and
(2) generating the output as a piece of edited code.
We show a model architecture as \autoref{fig:previous-work} to capture the general idea of the state-of-the-art solutions,
where optional edit description, prior edits and their optional context, and the code under the edit are fed
to a language model to output a piece of edited code.

While those solutions have laid an important foundation for code editing tasks,
there is still a gap between the solutions and the practical scenarios.
\begin{itemize}[leftmargin=*]
  \item \textbf{Assumption of Relevance of Prior Edits.}
    In an editing session, existing work usually assumes that all the prior edits of a target edit are relevant.
    However, it might not be true in practice (see Section~\ref{sec:example} for more details).
    Feeding the model with irrelevant prior edits can introduce noisy input,
    compromising the accuracy of the generated edits.
  \item \textbf{Assumption of Availability of Subsequent Edit Location.}
    In addition, it is also non-trivial to know where the edits can happen
    because the ripple effect of a prior edit may propagate across the whole project \cite{ufuktepe2022tracking}.
  \item \textbf{Interactive Nature between Multiple Edits.}
    Lastly, code edits can interact with each other regarding their
    syntactic dependency and
    semantic relevance.
    However, existing transformers still lack of design to capture such interaction.
\end{itemize}

In this work, we propose, \tool, an LM-based solution to address the above concerns.
We designed \tool to monitor the ripple effect of an edit as to where the subsequent edits can happen,
infer the most relevant prior edits, and
capture the edit interaction more explicitly.
To this end, we design \tool by
orchestrating a set of neural transformers \cite{vaswani2017attention} to coherently work with each other.
Once an edit-triggering event happens (e.g., an edit $e$ happens with an optional edit description $prp$), the following components are activated in an order:
\begin{itemize}[leftmargin=*]
  \item \textit{Two-staged edit location:}
    In the first stage, we scan the whole project with an \filelocator,
    which reports a set of files $\mathcal{F}$ where the changes can happen in a coarse-grained way.
    In the second stage, with the reported files $\mathcal{F}$, we apply a sliding window on those files
    with our \linelocator to report the type of edit (e.g., \textit{keep}, \textit{insert}, and \textit{replace}) for each line of code in the files.
    As a result, we can have a set of lines of code with labelled type of edit,
    denoted as $\mathcal{L}_e = \{l_e = (l, t) | l\in \mathcal{L}, t\in\{$\textit{insert}, \textit{replace}$\}\}$,
    where $\mathcal{L}$ is the set of lines of code in the project.
    $\mathcal{L}_e$ includes all the lines of code predicted to be inserted with or replaced with new content.
  \item \textit{Edit content generation:}
    With the reported editing locations $\mathcal{L}_t$,
    we use our trained \generator to further generate the edit content for each location with prediction $e_t = (l, t)$, regarding the editing description $prp$ and a set of selected relevant prior edits.
    Specifically, we select a set of relevant prior edits $\mathcal{P} = \{e = (l, t, c_a, c_b)\}$ to generate a list of edit options,
    where $l$ indicates the editing line of code, $t$ indicates the edit type, $c_a$ indicates the code content after the edit, and $c_b$ indicates the code content before the edit.
    Note that, $c_a$ and $c_b$ further incorporate user feedback on the code under the edit,
    allowing us to adapt the user's intention on-the-fly in the editing session.
  \item \textit{Edit-dependency analyzer:}
    For selecting the relevant prior edits, we train an \depanalyzer to parse all the prior edits and select the most syntactically and semantically relevant ones for generating the target edit.
\end{itemize}
Once a new edit $e'$ is accepted, it serves as a new edit-triggering event to activate the above procedures.

\definecolor{codegreen}{rgb}{0,0.6,0}
\definecolor{codered}{rgb}{0.6,0,0}
\definecolor{codegray}{rgb}{0.5,0.5,0.5}

\lstdefinestyle{lgeometry}{ xleftmargin = 20pt, xrightmargin = 0pt, frame = tb, framesep = \fboxsep, framexleftmargin = 20pt}

\lstdefinestyle{gostyle}{
    language=Go,
    basicstyle=\ttfamily\footnotesize,
    breakatwhitespace=false,
    breaklines=true,
    captionpos=b,
    keepspaces=true,
    showspaces=false,
    showstringspaces=false,
    showtabs=false,
    tabsize=2,
    commentstyle=\color{codegray},
    keywordstyle=\color{blue},
    stringstyle=\color{codegreen},
    moredelim=[is][\color{codered}\textbf]{@}{@},
    moredelim=[is][\color{codegreen}\textbf]{^}{^},
    columns=flexible,
    aboveskip=-0.5\baselineskip,
    belowskip=-0.5\baselineskip,
}
\lstset{style=gostyle}

\noindent
\begin{table*}[t]
\caption{
The code edits in src/testing/benchmark.go
}
\label{tab:example11}
\centering
\small
\vspace{-10pt}
\begin{tabular}{|p{0.1\linewidth}|p{0.4\linewidth}|p{0.4\linewidth}|}
\hline
\textbf{Hunk} & \textbf{\centering Before Edit} & \textbf{\centering After Edit} \\
\hline
H1 (insert) &
\begin{lstlisting}
type benchContext struct {


	maxLen int // The largest recorded benchmark name.
}
\end{lstlisting}
&
\begin{lstlisting}
type benchContext struct {
	^match *matcher^

	maxLen int // The largest recorded benchmark name.
}
\end{lstlisting}
\\
\hline

H2 (insert)
&
\begin{lstlisting}
func runBenchmarksInternal(...) bool {
	// ... other code ...
	ctx := &benchContext{


		extLen: len(benchmarkName("", maxprocs)),
	}
 	// ... other code ...
}
\end{lstlisting}
&
\begin{lstlisting}
func runBenchmarksInternal(...) bool {
	// ... other code ...
 	ctx := &benchContext{
		^match:  newMatcher(matchString, *matchBenchmarks, "-test.bench"),^
		extLen: len(benchmarkName("", maxprocs)),
	}
 	// ... other code ...
}
\end{lstlisting}
\\

\hline

H3 (replace)
&
\begin{lstlisting}
func (b *B) runBench(...) bool {
	// ... other code ...
	@if b.level > 0 {
		name = b.name + "/" + name@





	}
 	// ... other code ...
}
\end{lstlisting}
&
\begin{lstlisting}
func (b *B) runBench(...) bool {
	// ... other code ...
	^benchName, ok := b.name, true
	if b.context != nil {
		benchName, ok = b.context.match.fullName(&b.common, name)
	}
	if !ok {
		return true^
	}
 	// ... other code ...
}
\end{lstlisting}
\\
\hline

\end{tabular}
\end{table*}

\begin{table*}[t]
\caption{The code edits in src/testing/testing.go}
\label{tab:example12}
\small
\centering
\vspace{-10pt}
\begin{tabular}{|p{0.1\linewidth}|p{0.4\linewidth}|p{0.4\linewidth}|}
\hline
\textbf{Hunk} & \textbf{Before Edit} & \textbf{After Edit} \\
\hline

H4 (insert)
&
\begin{lstlisting}
type testContext struct {


	mu sync.Mutex
   //  ... other code ...
}
\end{lstlisting}
&
\begin{lstlisting}
type testContext struct {
	^match *matcher^

	mu sync.Mutex
   //  ... other code ...
}
\end{lstlisting}
\\
\hline
H5 (replace)
&
\begin{lstlisting}
func (t *T) run(...) bool {
	@testName := name
	if t.level > 0 {
		testName = t.name + "/" + name@

	}
  // ... other code ...
}
\end{lstlisting}
&
\begin{lstlisting}
func (t *T) run(...) bool {
	^testName, ok := t.context.match.fullName(&t.common, name)
	if !ok {
		return true^
	}
   // ... other code ...
}
\end{lstlisting}
\\

\hline

H6 (replace)
&
\begin{lstlisting}
@func newTestContext(maxParallel int) *testContext {@
	return &testContext{

		startParallel: make(chan bool),
		maxParallel:   maxParallel,
		running:       1, // Set the count to 1 for the main (sequential) test.
	}
}
\end{lstlisting}
&
\begin{lstlisting}
^func newTestContext(maxParallel int, m *matcher) *testContext {^
	return &testContext{
		^match:         m,^
		startParallel: make(chan bool),
		maxParallel:   maxParallel,
		running:       1, // Set the count to 1 for the main (sequential) test.
	}
}
\end{lstlisting}
\\
\hline

\end{tabular}

\end{table*} 

We train our neural models from the collected over 180K commits from 471 open source projects in 5 programming languages.
We evaluate our models with extensive experiments.
Our extensive experiment shows that
(1) \tool can identify edit locations with an accuracy of 70.8-85.3\%; and
(2) for each identified edit location, \tool achieves the exact match rate of 41.8\%
and the BLEU score of 60.7 for the top-1 recommendation.
Our ablation study shows that \tool, as a code-editing framework, can improve the exact match rate and BLEU score of state-of-the-art edit generators such as
GRACE and CoditT5 by on average 8.57\% and 18.08 respectively.
Further, our user study on 18 participants with 3 editing tasks on feature enhancement, refactoring, and bug fixing
shows that
 (1) in comparison to our baseline Copilot,
\tool can be effective in assisting users to edit code by its advantage
on the project-wise awareness and the capture of the interaction between relevant edits, and
(2) sheds light on the future improvement of the tool design such as distribution-shifting edits from the training dataset.

Overall, we summarize our contributions as follows:
\begin{itemize}[leftmargin=*]
  \item We propose \tool, an LM-driven solution to make the state-of-the-art edit generation models more practical by predicting the relevant prior edits, subsequent edit location, and the interactive nature between the edits.
  \item We design \tool as a modularized framework, which allows us to plug into any edit-content generators in the community.
  \item We implement an open-source \tool as a VS Code plugin,
    which adopts a cloud infrastructure and allows the programmers to try in practice with convenience.
  \item We conduct extensive experiments (simulation, model-wise evaluation, and user study) showing the effectiveness of individual models as independent model design, model interaction as a whole system, and UI design as a tool.
\end{itemize}

Given the space limit, the tool video demonstration, experimental details, and further discussion are available at \cite{code-edit-pilot}. 
\section{Motivating Example}\label{sec:example}

\autoref{tab:example11} and \autoref{tab:example12} (implemented by Go programming language) shows a simplified code-editing example from the commit \texttt{00a2} under the project \texttt{golang/go}\footnote{The address can be referred in https://github.com/golang/go}.
We summarize the programmer's editing intention in such a commit as follows.

\noindent\textbf{Original Design.}
The function under edit is to update the way to select test cases and benchmark in the \texttt{golang/go} project.
The Go project is delivered with the \texttt{testing} package where
a set of test cases are used to check the performance on a set of benchmarks of Go programs.
The source file \texttt{src/testing/testing.go} automates the testing of the project by selecting a subset of required test cases, and
the source file \texttt{src/testing/benchmark.go} selects a subset of the benchmark of Go programs such as runtime overhead, memory allocation, lock performance, etc.
The old implementation of selecting test cases and benchmarks is by keyword-based matching the name of benchmark and test suites with a string (see the hunk of \textit{Before Edit} of H3 in \autoref{tab:example11} and H5 in \autoref{tab:example12}).

\noindent\textbf{Editing Intention.}
In the editing session,
the programmer intended to introduce a regular expression matcher to select the required
benchmark and test cases.

\noindent\textbf{Editing Implementation.}
To this end, the programmer edits the \texttt{benchmark.go} and \texttt{testing.go} files as follows:
\begin{itemize}[leftmargin=*]
  \item \textbf{H1} (see \autoref{tab:example11}):
    Introduce a variable \texttt{matcher} with pointer type \texttt{*matcher} in the type \texttt{benchContext};
  \item \textbf{H2} (see \autoref{tab:example11}):
    Introduce a parameter of type \texttt{matcher} when initializing a reference of \texttt{benchContext};
  \item \textbf{H3} (see \autoref{tab:example11}):
    Replace the keyword-based matching implementation with regular-expression-based matching implementation;
  \item \textbf{H4} (see \autoref{tab:example12}):
    Introduce a variable \texttt{matcher} with pointer type \texttt{*matcher} in the type \texttt{testContext};
  \item \textbf{H5} (see \autoref{tab:example12}):
    Replace the keyword-based matching implementation with regular-expression-based matcher implementation;
  \item \textbf{H6} (see \autoref{tab:example12}):
    Introduce a parameter of type \texttt{matcher} when initializing a reference of \texttt{testContext};
\end{itemize}

\begin{figure}[t]
  \centering
  \includegraphics[scale=0.4]{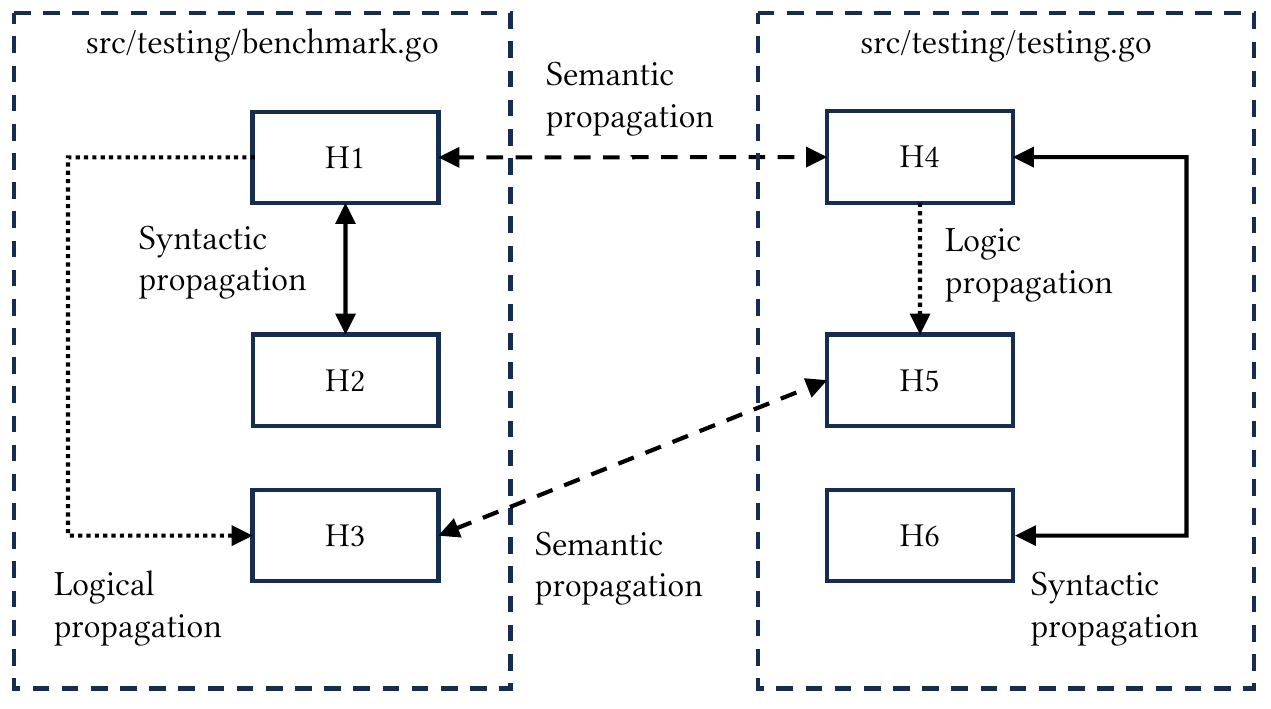}
  \caption{The type of edit propagation in the code-editing example showed in \autoref{tab:example11} and \autoref{tab:example12}.}
  \label{fig:motivating-example}
  \vspace{-10pt}
\end{figure}

While each edit is a simple operation, they are interactive and relevant as different types of edit propagation as shown in \autoref{fig:motivating-example}.
Following the notation in \autoref{tab:example11} and \autoref{tab:example12},
we use H$_i$ ($i$ =  1, ..., 6) to indicate the hunk in the code example.

\begin{itemize}[leftmargin=*]
  \item \textbf{Syntactic Propagation:}
    Syntactic propagation indicates that an edit $e_i$ incurs a compilation error in the project,
    which further mandates another edit $e_j$ to fix the error.
    For example, hunk H1 happens so as to cause a compilation error on the location of hunk H2 for missing an initialized parameter.
    In \autoref{fig:motivating-example}, the fact that H1 and H2 point to each other indicates that the edit propagation caused by program syntax is mutual.
  \item \textbf{Semantic Propagation:}
    Semantic propagation indicates that an edit $e_i$ is propagated to $e_j$ because $e_i$ and $e_j$ are applied to similar functionalities.
    In \autoref{fig:motivating-example}, for the example of the editing pair (H1, H4) and (H3, H5),
    an edit can propagate to the other edit in the pair.
  \item \textbf{Logical Propagation:}
    Logical propagation indicates that an edit $e_i$ lays a foundation for another edit $e_j$ to accomplish a task.
    In \autoref{fig:motivating-example}, H1 does not necessarily cause a compilation error at H3, however, H1 introduces a variable \texttt{matcher} so that the matching implementation is updated at H3.
\end{itemize}

Thus, we can see that
(1) the edits are interactive with each other in a different way,
(2) only a limited number of prior edits is relevant and informative to contribute to an edit, and
(3) the edit can propagate to any possible files in the project.
However, despite that the existing state-of-the-art solutions such as GRACE \cite{GRACE}, CCT5 \cite{lin2023cct5}, MODIT \cite{MODIT} and CoditT5 \cite{ZhangETAL22CoditT5} lay an important foundation (see the summary their model architecture in \autoref{fig:previous-work}),
they are still far from accomplishing the edit recommendation tasks in the aforementioned practice.

\section{Approach}

\begin{figure}[t]
  \centering
  \includegraphics[scale=0.38]{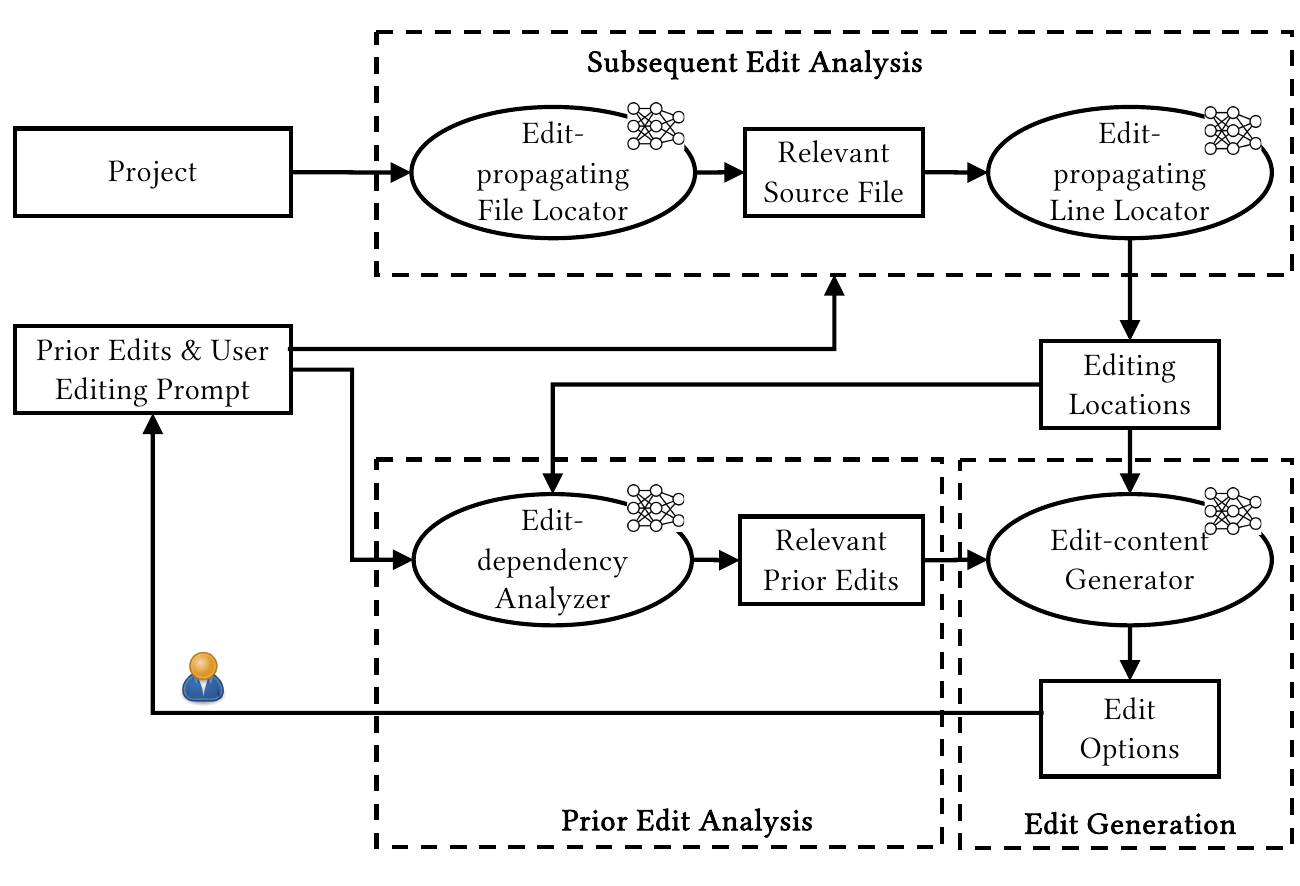}
  \caption{Overview of \tool, consisting of subsequent edit analysis, edit generation, and prior edit analysis.
  The analysis is triggered once an edit-trigger event happens.
  \tool orchestrates a set of neural-transformer-based components to accomplish the editing task}
  \label{fig:overview}
  \vspace{-10pt}

\end{figure}

\autoref{fig:overview} shows an overview of our \tool design,
which takes a set of prior edits and an optional edit prompt, and
generates the output as a list of subsequent editing locations and their editing options.
Overall, the \tool architecture consists of subsequent edit analysis, prior edit analysis, and edit generation.

\begin{itemize}[leftmargin=*]
  \item \textbf{Subsequent Edit Analysis} takes a set of selected prior code edits and an optional editing prompt to estimate the subsequent edits in the project.
      In this work, we adopt a two-stage estimation.
      The first stage estimates the relevant source files, with \filelocator, for where the subsequent edits can happen in a coarse-grained (and lighted) way.
      The second stage further applies a fine-grained detector (i.e., \linelocator) to predict the editing type of each line of code in those files.
  \item \textbf{Prior Edit Analysis} takes the editing locations and selects the most relevant prior edits with \depanalyzer, regarding their potential of syntactic, semantic, and logical edit propagation to a target edit location.
  \item \textbf{Edit Generation} generates the concrete edit options for each edit location with predicted editing type of \textit{insert} and \textit{replace}, regarding the selected prior edits.
      Note that, once the user confirms a recommended edit option by
      (1) directly accepting our recommendation,
      (2) modifying based on our recommendation, or
      (3) input his or her own edit,
      it will be included as a new prior edit.
      Further, the newly applied edit serves as a new edit-triggering event to launch a new round of editing recommendations.
\end{itemize}


\subsection{Subsequent Edit Analysis}

\noindent\textbf{Problem Statement.}
We consider the problem of finding the subsequent edits with an edit and its optional user prompt as a problem of edit propagation.
Thus, we rephrase the problem as follows.
Given a project be a set of files $P$, the user's editing prompt be $prp$,
the latest edit $e = (c_b, c_a)$ where $c_b$ is the code before edit, and $c_a$ is the code after edit,
we aim to locate a subset of files $F \subset P$,
where each $f\in F$ specifies the subsequent edits by attaching each line of code with an editing type as \textit{keep}, \textit{insert}, or \textit{replace}.

\noindent\textbf{Challenge.}
As mentioned above, the edits can interact with each other regarding the syntactic dependency and semantic relevance.
As for analyzing syntactic dependencies,
we usually need to parse the whole compilable project to build the program dependency graph \cite{ferrante1987program} to track the data, control, and call dependencies.
However, the graph construction for large projects could be time-consuming.
Further, the implementation of syntactic graph construction \cite{vallee2010soot, ali2013averroes} and semantic relevance \cite{keivanloo2015threshold, duncan2021pyclone, alfageh2020clone} are usually language-dependent.
Therefore, we use the neural models for estimating both the syntactic dependency and semantic relevance between two pieces of source code in a more runtime-efficient and language-independent way.

In this work, we adopt a two-stage localization solution, i.e.,
file localization in a coarse-grained way and line of code localization in a fine-grained way.

\subsubsection{Propagation File Localization}\label{sec:file-propagation}

Technically, we select a subset $F' \subset F$ where $F' = \{f| sub_{edt}(f, e) > th_{sub}, f\in F\}$,
where $sub_{edt}(., .)$ is a likelihood estimation function for the file $f$
which can be co-edited given the input edit $e$.
Further, $th_{sub}$ is a threshold to estimate its likelihood.

We estimate the propagation likelihood regarding two factors, i.e.,
(1) the estimated dependency of the file $f$ on $e$, and
(2) the semantic similarity between some code in $f$ and $e$.
Namely, we design \autoref{eq:propagation-file} as follows.
\begin{equation}\label{eq:propagation-file}
sub_{edt}(f, e) = \alpha_1\cdot dep(e, f) + \alpha_2\cdot sem(e, f) + \epsilon
\end{equation}

In \autoref{eq:propagation-file}, we let
each coefficient $\alpha_i > 0$.
We quantize each factor (estimated dependency $dep(e, f)$ and semantic similarity $sem(e, f)$)
as a score between 0 and 1 as follows.

\begin{figure}[t]
  \centering
  \includegraphics[scale=0.5]{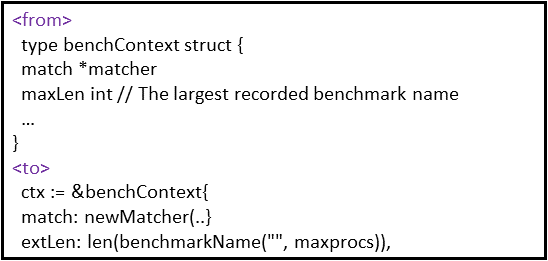}
  \caption{An example of input of our transformer for learning the dependency.}
  \label{fig:dependency-input}
  \vspace{-10pt}
\end{figure}

\paragraph{Estimated Dependency}\label{sec:estimated-dependency}
Given an edit $e=(c_b, c_a)$ and a source file $f$, we develop a dependency inference function $dep(e, f)$ to quantize the likelihood that $f$ depends on $e$.
Technically, we use a transformer (e.g., CodeT5 and CodeBERT) as our base model to learn the dependency between the source code.
We follow the design of GRACE \cite{GRACE} by constructing the input of a transformer-based language model as shown in \autoref{fig:dependency-input}.
Specifically, we use the tags \texttt{<from>} and \texttt{<to>} as the separator between two pieces of source code.
Those tags play a role as instruction tuning.
Then we add one dense layer to have two output neurons activated with sigmoid function, i.e.,
(1) the former code depends on the latter code and
(2) the latter code depends on the former code.
Given a pair of source code $c_1$, $c_2$, their labelled dependencies are $y_1$ and $y_2$ ($y_1 = 1$ or 0 is for whether $c_1$ depends on $c_2$ and $y_2 = 1$ or 0 is for whether $c_2$ depends $c_1$),
and their estimated dependency are $\hat{y_1}$ and $\hat{y_2}$,
we design the loss function as \autoref{eq:loss-dependency}:
\begin{equation}\label{eq:loss-dependency}
    \begin{split}
       loss(c_1, c_2) & = - (y_1\times log(\hat{y_1}) + (1-y_1)\times log(1 - \hat{y_1}) +\\
         & y_2\times log(\hat{y_2}) + (1-y_2)\times log(1-\hat{y_2}))
    \end{split}
\end{equation}

In this work, we use Jin et al.'s dependency analyzer \cite{8802634, 9765666} to construct the training dataset.
Limited by the input length,
we split a file $f$ into $k$ smaller segments as $seg_1, .., seg_k$.
Further, we choose $c_b$ (the code before the edit) of the latest edit as the target code $c_{tar}$.
Then we estimate the likelihood of the dependency between $c_{tar}$ and each code segment.
For convenience, we use the symbol of the second output neuron $\hat{y_2}$ to denote the likelihood of the latter code to depend on the former code,
$dep(e, f) = max(\hat{y_2}(c_{tar}, seg_i))$.
That is, we adopt one-directional dependency to infer the edit propagation.
Further, we select the $max(.)$ function as we favour the recall over the precision in this stage.
By replacing the analyzer tool \cite{8802634, 9765666} with a neural network, we reduce the runtime overhead of analyzing a pair of code snippets from $\sim$70 seconds to $\sim$0.01 second.

\paragraph{Semantic Similarity and Prompt Relevance}\label{sec:clone}

We capture the semantic similarity of code-to-code by neural embedding in a universal way.
The rationale is that we believe a pretrained neural network such as CodeT5 and CodeBERT can capture both the syntactic and semantic similarity.
Therefore, still considering the limit of input length of a transformer,
we split a source file $f$ into $k$ segments as $seg_1, ..., seg_k$, $c_{tar} = c_{b}$ where $c_b$ is the code before the edit,
and $emd(.)$ as the representation of a piece of code or a prompt extracted from the transformer,
we can have:
\begin{equation}\label{eq:code2code}
  sem(e, f) = max(cos(emd(c_{tar}), emd(seg_i)))
\end{equation}

By this means, with given hyperparameters $\alpha_1$, $\alpha_2$, $\epsilon$, and $th_{sub}$,
we have a set of reported source files in a coarse-grained way. These coefficients, intercept and thresholds are available at \cite{code-edit-pilot}.

\subsubsection{Propagation Line Localization}\label{sec:propagation-line}
\begin{figure}[t]
  \centering
  \includegraphics[scale=0.43]{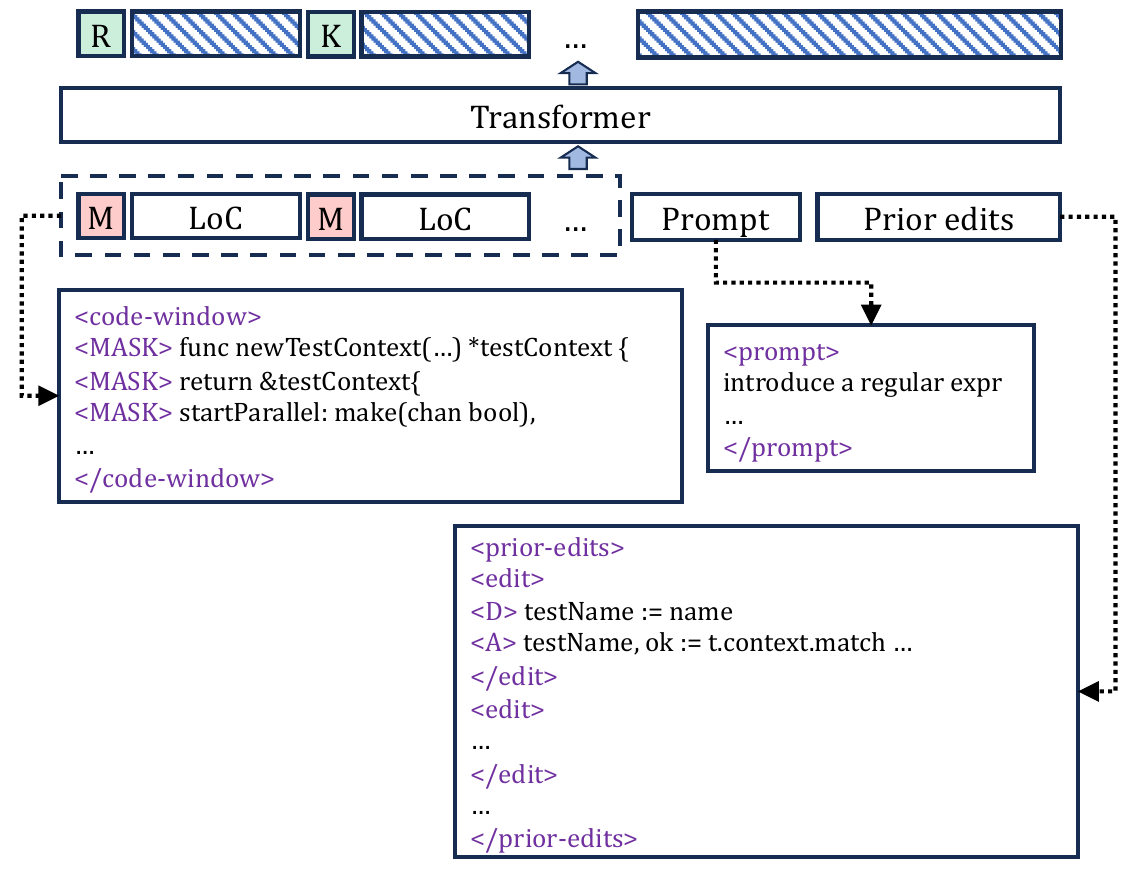}
  \caption{Overview of fine-grained edit location architecture.
  We formulate the edit location problem as a MLM (Mask Language Modelling) problem to predict the edit type of each LoC (Line of Code).}
  \label{fig:edit-locator}
  \vspace{-10pt}
\end{figure}

Given the located source files with the propagation potential,
we apply a sliding window across each file to identify the editing type of each line of source code.
As shown in \autoref{fig:edit-locator}, we fine-tune a base transformer model as a MLM (Mask Language Modeling) \cite{devlin2018bert} problem by instruction tuning \cite{salazar2019masked}.
Overall, the input of the transformer consists of the target code inside the window, the user prompt, and the relevant prior edits (see more details in Section~\ref{sec:prior-edit-analysis}).
For each input component, we introduce instructions (or tags) such as
\texttt{code-window}, \texttt{prompt}, \texttt{prior-edits} and \texttt{edit} as the separators
for the model to learn the input structure.
For each line of the code, we additionally introduce an operator as follows:
\begin{itemize}[leftmargin=*]
  \item \textit{keep}: the operator type indicates that a line should not be changed, symbolled as \texttt{<K>}.
  \item \textit{insert}: the operator type indicates that there shall be some code inserted after the line, symbolled as \texttt{<I>}.
  \item \textit{replace}: the operator type indicates that the line should be replaced by either an empty line (i.e., delete) or a few different lines (update), symbolled as \texttt{<R>}.
\end{itemize}
These edit operators are masked with a special token \texttt{<MASK>} in input.
Therefore, we apply MLM task on the operators to train the model to recover them.
The prompt is collected from the commit message from the code commit histories.
Further, we introduce the details of selecting the prior edits
in Section~\ref{sec:prior-edit-analysis} and Section~\ref{sec:model-training}.

\subsection{Prior Edit Analysis}\label{sec:prior-edit-analysis}

\noindent\textbf{Problem Statement.}
Given a set of prior edits $E_p$ = $\{e_{p_1}, ..., e_{p_k}\}$ where $e_i = (c_{b_i}, c_{a_i})$,
and the target code $c_{b_{tar}}$,
we quantize the likelihood of the influence of $e_{p_i}$ to $c_{b_{tar}}$ between 0 and 1.
Specifically, we denote the estimation function as $rel(.,.): E_p \times C \rightarrow (0, 1)$,
where $C$ is the set of pieces of code,
i.e., $rel(e_i, c_{b_{tar}}) \in (0, 1)$.

We quantize the relevance of prior edits by their syntactic dependency and semantic similarity
by \autoref{eq:prior-edit}:

\begin{equation}\label{eq:prior-edit}
    \begin{split}
        rel(e_{p_i}, c_{b_{tar}}) & = FCN(dep(e_{p_i}, c_{b_{tar}}), sem(e_{p_i}, c_{b_{tar}}), \\
        & loc_{sim}(e_{p_i}, c_{b_{tar}}))
    \end{split}
\end{equation}

Further, in \autoref{eq:prior-edit}, $FCN$ is a multi-layer fully connected network, the dependency estimation function $dep(.,.)$ for estimating the dependency from $c_{c_{tar}}$ to the code before the edit of $e_{p_i}$ and the semantic relevance function $sem(.,.)$ is defined in Section~\ref{sec:file-propagation}.
Function $loc_{sim}$ evaluates the proximity between $e_{p_i}$ and $c_{b_{tar}}$ as:
\begin{equation}\label{eq:locality}
  loc_{sim}(e_{p_i}, c_{b_{tar}}) =
  \begin{cases}
    1- \frac{|loc(e_{p_i}) - loc(c_{b_{tar}})|}{k} & \text{if } ld(e_{p_i},c_{b_{tar}})<k \\
    0 & otherwise
    \end{cases}
\end{equation}
In \autoref{eq:locality}, we use a sliding window of size $k$ to define whether the location difference of $e_{p_i}$ and $c_{b_{tar}}$ is small (i.e., $ld(e_{p_i},c_{b_{tar}})<k$).
If it is, we estimate the proximity as \autoref{eq:locality}.
Otherwise, the function $loc_{sim}(., .)$ is 0.
Finally, we define a threshold $th_{pri}$ to identify the set of relevant prior edits $E_{rel} = \{e_{p} | rel(e_p, c_{b_{tar}}) > th_{pri}\}$.


\subsection{Edit Generation}\label{sec:edit-generation}

\begin{figure}[t]
  \centering
  \includegraphics[scale=0.39]{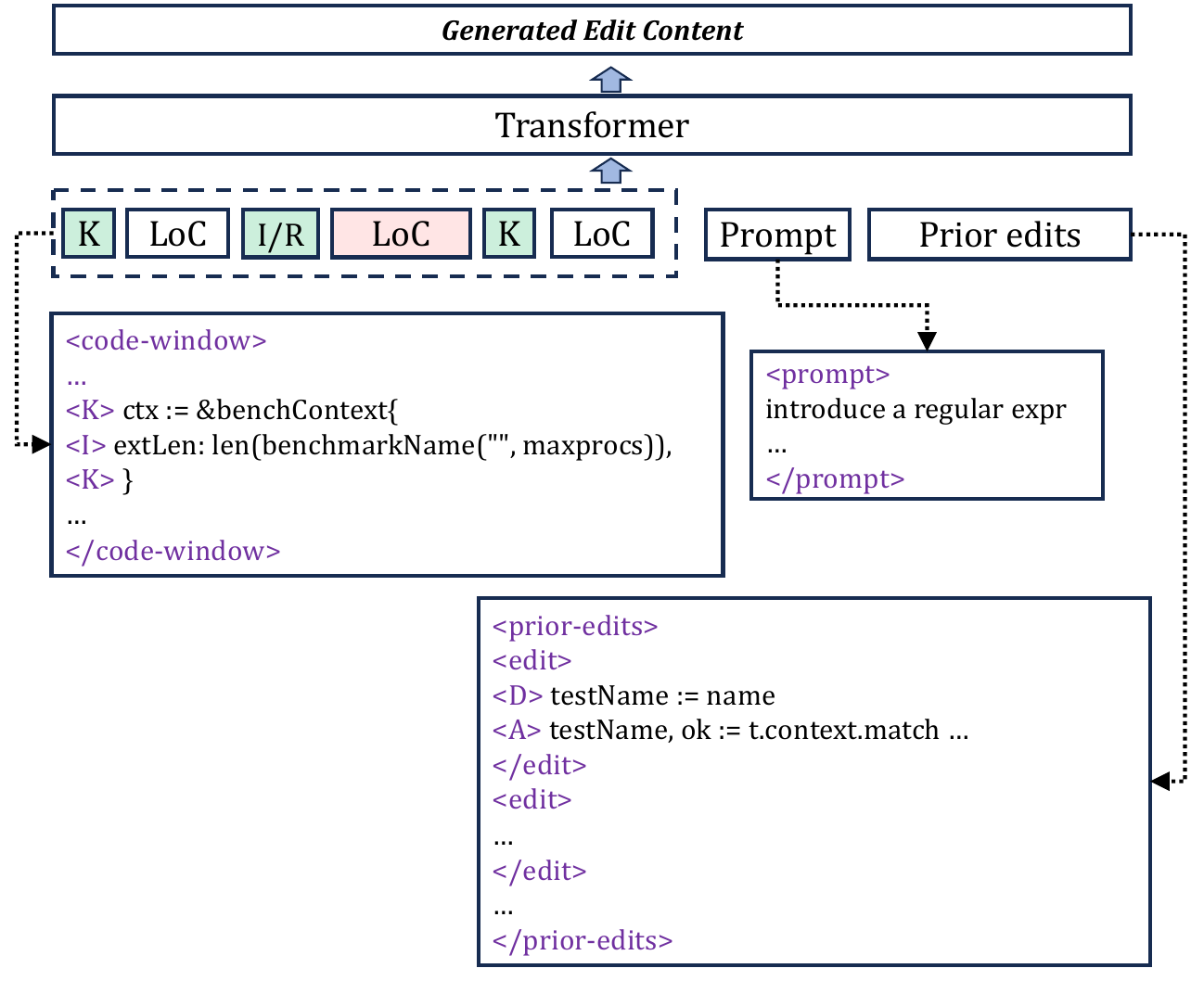}
  \caption{Overview of edit generator, which generates the edit content for a fine-grained edit location.}
  \label{fig:edit-generation}
  \vspace{-10pt}

\end{figure}

\autoref{fig:edit-generation} shows the overall model architecture to generate edit content on \textit{one} edit location and selected prior edits.
Similar to the design of the location of edit lines,
the edit generation model takes as input
a code-window under the edit,
the user's prompt, and
relevant prior edits.
The prompt and the prior edits share similar tags for the model to capture the structure.

In contrast, the code window describes a \textit{hunk} consists of consecutive lines of the same edit type (\textit{replace} and \textit{insert}) with a few lines of type \textit{keep} as its context.
Specifically, each line is attached with a tag \texttt{<K>} for the edit type of \textit{keep};
and with a tag \texttt{<I>} or \texttt{<R>} for the edit type of \textit{insert} and \textit{replace} respectively.
Further, the output predicts the edit content of the edit location.
We train the transformer with classical cross-entropy loss \cite{CrossEntropyLoss}.
On the runtime, we use Beam Search \cite{freitag2017beam} to generate $k$ edit options ranked with their confidence.
Last but not least, the user can either accept or modify upon our recommended edits,
the new edit will be kept as a new prior edit as user feedback,
to further facilitate the whole editing session.

\subsection{Model Training}\label{sec:model-training}

Overall, we have the following neural models to train, i.e.,
an \depanalyzer (see Section~\ref{sec:estimated-dependency}),
an \linelocator (see Section~\ref{sec:propagation-line}), and
an \generator (see Section~\ref{sec:edit-generation}).

We first train the \depanalyzer by collecting the dependency of source code by running Jin et al.'s dependency analysis tool \cite{8802634, 9765666} on the open-source projects.
Note that, our neural dependency analyzer is expected to predict the dependency between arbitrary two pieces of code without the awareness of their programming language.
Then, we train the \linelocator and an \generator in an interactive manner.
We collect the commits from the open source projects as the training dataset (see Section~\ref{sec:experiment}).
For each commit, we take one hunk as an individual edit,
then we train our models by estimating the random order of both intra-file edits and inter-file edits.
The rationale is that we do not know the sequence of files being edited and that of the edited locations within a file.
Therefore, we do not make any editing partial order assumption on those edits.

Further, given a set of prior edits,
we normalize their relevance into a probability distribution $X$.
For example, assume that we have three prior edits with the relevance to an edit and a prompt (i.e., the editing description) as 0.7, 0.3, 0.6,
then we normalize their relevance to $X = \{\frac{0.7}{0.7+0.3+0.6}, \frac{0.3}{0.7+0.3+0.6}, \frac{0.6}{0.7+0.3+0.6}\}$
= $\{0.437, 0.187, 0.375\}$ to sample the prior edits during the training.

\section{Tool Design}\label{sec:tool}
\begin{figure*}[t]
  \centering
  \includegraphics[scale=0.63]{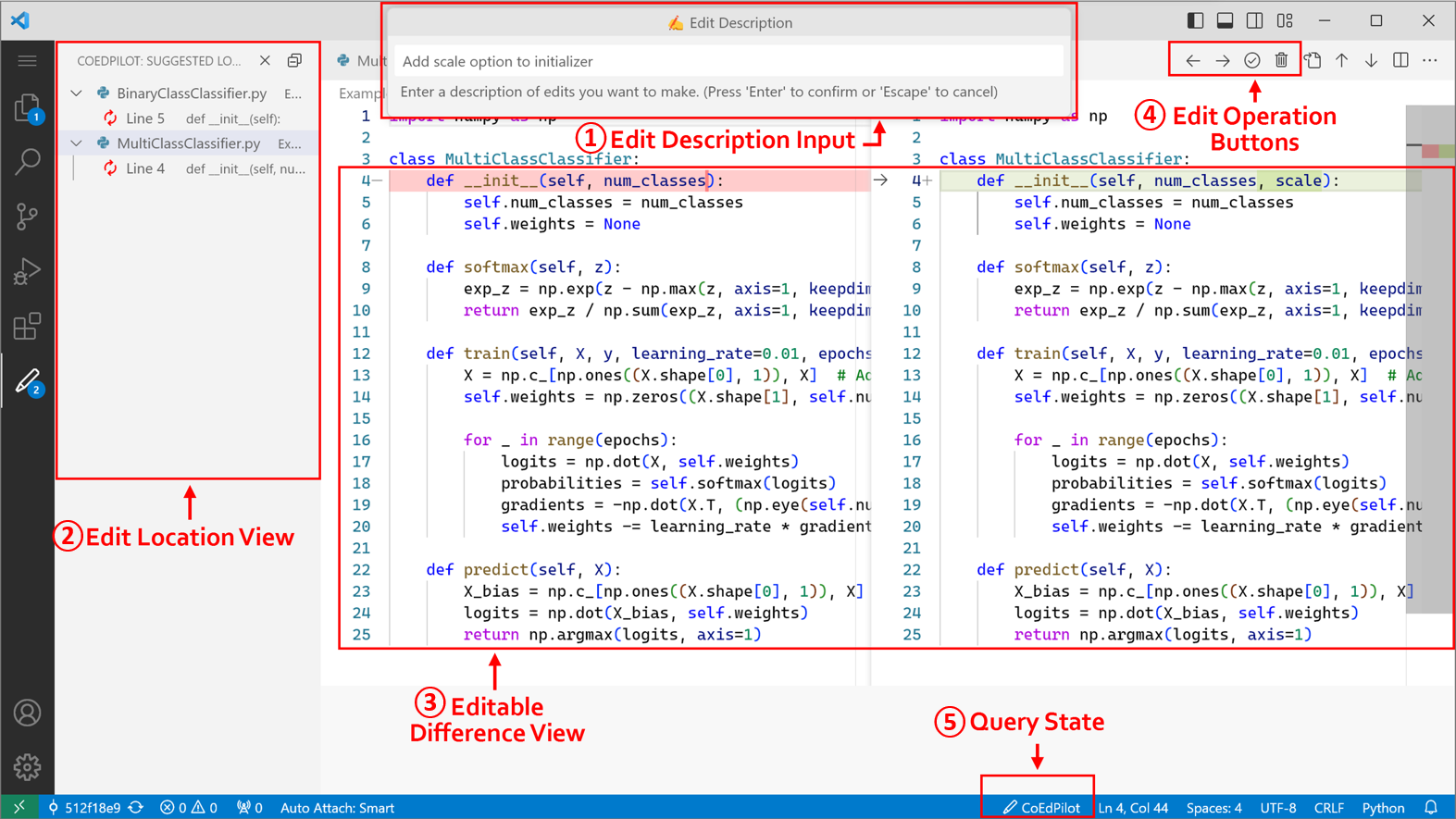}
  \caption{The screenshot of \tool tool, implemented as a Visual Studio Code extension.}
  \vspace{-5pt}
  \label{fig:tool}
\end{figure*}

\autoref{fig:tool} shows a screenshot of our \tool tool as a Visual Studio Code extension \cite{vscode}, which consists of functions designed according to our approach (see \autoref{fig:overview}).
We introduce the basic functions and GUI (see \autoref{fig:tool}) as follows. 
A detailed video is available at \cite{code-edit-pilot}.
\begin{itemize}[leftmargin=*]
  \item \textbf{Triggering the Edit Recommendation}:
    When the users edit the code, they can trigger the edit recommendation with a shortcut (or right-click the editor) to request edit locations.
    Then, an \textit{Edit Description Input} \ding{172} will be shown for them to input their optional description of the edit.
  \item \textbf{Subsequent Edit Recommendation}:
    Then, \tool shows an \textit{Edit Location View} \ding{173} where the edit locations are organized in terms of edit files as their parent nodes and edit lines as their child nodes.
    The users can click a child node to highlight the corresponding location in the code editor,
    where a line with edit type of \textit{insert} is in green and a line with that of \textit{replace} is in red.
  \item \textbf{Edit Option Recommendation}:
    Next, the users can further request the edit option in each edit location, as shown in \textit{Editable Difference View} \ding{174} in \autoref{fig:tool} where how the code before and after the edit is simulated.
    Users can use the \textit{Edit Operation Button} \ding{175} to \textit{browse},  \textit{accept} and \textit{ignore} the edit options.
    The accepted edit (and their follow-up modification) will be recorded as prior edits for next recommendation.
  \item \textbf{Cloud Service}:
    Last, we follow the design of Copilot to deploy \tool on the cloud so that the user request (e.g., for edit location and edit generation) and their response are communicated between the server and the client. Users can check the network connection by \textit{Query State} \ding{176} as in \autoref{fig:tool}.
\end{itemize} 
\section{Experiment}\label{sec:experiment}

We evaluate \tool with the following research questions:

\captionsetup[table]{skip=8pt}
\begin{itemize}[leftmargin=*]
  \item \textbf{RQ1 (Locating Propagating Files, see Section~\ref{sec:file-propagation})}: Can \tool locate the edit-propagating source files?
  \item \textbf{RQ2 (Locating Propagating Lines, see Section~\ref{sec:propagation-line})}: Given the located source files, can \tool locate the edit-propagating lines of code?
  \item \textbf{RQ3 (Edit Generation, see Section~\ref{sec:edit-generation})}: Given edit location, what is the performance of generating edit options?
  \item \textbf{RQ4 (Prior Edit Relevance, see Section~\ref{sec:propagation-line})}: Can \tool select the relevant prior edits accurately?
  \item \textbf{RQ5 (Performance Boost for State-of-the-arts)}:
    Whether the framework of \tool further boost the performance of the state-of-the-art solutions?
\end{itemize}

Note that, \tool serves more as a complementary framework to enhance the state-of-the-art edit generator by locating subsequent edits and capturing relevant prior edits.
Thus, in RQ5, we briefly compare our edit generation model with the state-of-the-arts,
followed by how we can boost their performance.


\subsection{Benchmark Construction}
To evaluate the performance of \tool,
we construct a benchmark of 5 programming languages (i.e., JavaScript, Java, Go, Python, and TypeScript) from 471 open-source projects.
Upon construction, we select the top 100 projects from GitHub according to the number of their stars.
For each programming language, we remove the projects with educational purposes (e.g., tutorial) or non-English commit messages.
For each project,
we select commits with the following criteria\footnote{In this work, we provide our definition of good quality, but we encourage the practitioners to adjust the definition according to their practical scenarios.} in our dataset:
\begin{itemize}[leftmargin=*]
  \item A commit shall include at least three hunks;
  \item A commit shall include hunks with the number of changed lines of code less than 15 (considering the length limit of our model);
  \item The commit message shall be an English message with a token length over 5;
  \item The commit shall not contain the automatically generated source files (e.g., the Java files with \texttt{@auto} keywords) or non-source files (e.g., \texttt{.bak}, \texttt{.log}, and \texttt{.pyc} files)
\end{itemize}
As a result, we have the dataset as shown in \autoref{tab:benchmark}, with an average commit filter rate of 6.89\%.
Further, we train our dependency analyzer on 49 sampled projects of different programming languages with 77K positive pairs and 24K randomly sampled negative pairs.

\begin{table}
\centering
\caption{Benchmark of \tool on 471 open source projects on 5 programming languages.
For the columns of `Train', `Valid', `Test' dataset, we show the number of their training samples.}
\label{tab:benchmark}
\small
\tabcolsep=0.03cm
\begin{tabular}{ccccccccc}
\hline
\textbf{Language} & \textbf{Model} & \textbf{Train} & \textbf{Valid} & \textbf{Test} & \textbf{\#Proj} & \textbf{\#Com} & \textbf{\#File} & \textbf{\#Hunk} \\
\hline
\multirow{3}{*}{JavaScript} & File location & 22K & 3K & 6K & \multirow{3}{*}{93} & \multirow{3}{*}{34K} & \multirow{3}{*}{34K} & \multirow{3}{*}{658K} \\
\cline{2-5}
 & Line location & 382K & 54K & 109K &  &  &  &  \\
\cline{2-5}
 & Edit generation & 460K & 65K & 130K &  &  &  &  \\
\hline
\multirow{3}{*}{Java} & File location & 68K & 10K & 20K & \multirow{3}{*}{89} & \multirow{3}{*}{24K} & \multirow{3}{*}{72K} & \multirow{3}{*}{556K} \\
\cline{2-5}
 & Line location & 335K & 47K & 95K &  &  &  &  \\
\cline{2-5}
 & Edit generation & 389K & 55K & 111K &  &  &  &  \\
\hline
\multirow{3}{*}{Go} & File location & 46K & 7K & 14K & \multirow{3}{*}{98} & \multirow{3}{*}{50K} & \multirow{3}{*}{88K} & \multirow{3}{*}{1174K} \\
\cline{2-5}
 & Line location & 695K & 99K & 198K &  &  &  &  \\
\cline{2-5}
 & Edit generation & 822K & 117K & 234K &  &  &  &  \\
\hline
\multirow{3}{*}{Python} & File location & 60K & 9K & 17K & \multirow{3}{*}{91} & \multirow{3}{*}{33K} & \multirow{3}{*}{42K} & \multirow{3}{*}{555K} \\
\cline{2-5}
 & Line location & 327K & 46K & 93K &  &  &  &  \\
\cline{2-5}
 & edit generation & 389K & 55K & 111K &  &  &  &  \\
\hline
\multirow{3}{*}{TypeScript} & File location & 65K & 9K & 17K & \multirow{3}{*}{100} & \multirow{3}{*}{39K} & \multirow{3}{*}{76K} & \multirow{3}{*}{817K} \\
\cline{2-5}
 & Line location & 480K & 68K & 137K &  &  &  &  \\
\cline{2-5}
 & Edit generation & 572K & 81K & 163K &  &  &  &  \\
\hline
\end{tabular}
\vspace{-5pt}
\end{table}

\subsection{Experiment Setup}
\subsubsection{RQ1 (Propagating-file Location)}\label{sec:rq1-setup}
We extract a commit with $k$ hunks as a set, denoted as $H$, located in $m$ source files, to construct $k$ training samples.
In each sample, we select one hunk $h\in H$ as the target hunk.
Assume that the $m'$ files with other hunks as the ground-truth positive files, and we randomly select $n$ ($n > m$) files not in the commit as negative files.
If \tool reports $g$ files as positive and $h$ out of $g$ files are true positive,
we measure the precision of file location as $\frac{h}{g}$ and the recall of file location as $\frac{h}{m}$.

\subsubsection{RQ2 (Propagating-line Location)}\label{sec:rq2-setup}
We parse a commit with $k$ hunks as a set, denoted as $H$, located in $m$ files to construct $k$ training samples as follows.
Each time, we select one out of $k$ hunks, i.e., $h\in H$, as the target edit
to be predicted.
We select relevant prior edits from $H \setminus \{h\}$ with \tool.
Then we apply a sliding window of size $s$ across the $m$ files for \tool to report the hunk $h$.
We apply the above procedure for $k$ times, each of which we select a different target edit.

We measure the average accuracy, precision, and recall in the $k$ times as follows.
In each time, for the $m$ lines of code not in the prior edits,
we measure the accuracy as $\frac{n}{m}$,
where $n$ is the number of lines with edit type predicted accurately.
Further, we compute the precision and recall for each of the three labels individually.
Assume for each edit type, there are $l$ positive lines of code and \tool reports $t$ lines of code as positive and
$d$ out of $t$ lines are the true positive,
thus we have the precision as $\frac{d}{t}$ and the recall as $\frac{d}{l}$.
Given the imbalance in sample sizes across these labels, we employ the macro-averaging method to calculate the final precision and recall.

\subsubsection{RQ3 (Edit Generation)}\label{sec:rq3-setup}
Given a commit with the set of hunks as $H$, we then choose
one hunk $h$ as the target edit, and
have $H' = H \setminus \{h\}$, as the prior edits.
We use Beam Search to generate the top-1, top-3, top-5, and top-10 edit options for each edit location.
For each configuration, we measure its performance with
(1) the exact match rate (EMR) for a commit (i.e., an edit session) and
(2) the BLEU4 score \cite{papineni-etal-2002-bleu} of the generated edit content.
Specifically, assume that we generate the edit content exactly the same as the ground truth edit for $m$ out of $k$ times, the exact match rate is $\frac{m}{k}$.
Further, we calculate the highest BLEU4 score from all $k$ times' predictions.

\subsubsection{RQ4 (Prior Edit Prediction)}\label{sec:rq4-setup}
We compare training the edit locating models and the edit generation models with selective prior edits (by our \depanalyzer) and random prior edits.
We compare their performance as mentioned in Section~\ref{sec:rq2-setup} and Section~\ref{sec:rq3-setup}.

\subsubsection{RQ5 (Performance Boost)}\label{sec:rq5-setup}
We design the experiment as follows.
We select the state-of-the-art solutions, i.e., GRACE \cite{GRACE}, CCT5 \cite{lin2023cct5}, and CoditT5 \cite{ZhangETAL22CoditT5}, as the baselines, observe the boosting effect of \tool. 
CoPilot \cite{copilot} is neglected for its programming API is yet published at the time of this work.
\begin{itemize}[leftmargin=*]
  \item \textbf{Rough Edit Location} We provide the baselines with rough location as a general hunk area to observe their performance in generating edits.
  \item \textbf{Precise Edit Location} We equip baselines with our edit location model so that they are fed with specific lines to further observe their performance.
\end{itemize}

We measure the performance of edit generation as in Section~\ref{sec:rq5-setup}.
Given the space limit, we provide more experimental details (e.g., hyperparameters, hardware configuration, etc.) in our websites \cite{code-edit-pilot}.

\subsection{Results}\label{sec:results}

\subsubsection{RQ1 and RQ2 (Propagating-file Location \& Line)}\label{sec:rq1-res}

\autoref{tab:propagating-file} shows the overall performance of \tool to detect the edit locations regarding different granularity (i.e., file-level and line-level).
We achieve a average precision of 79.52\% and a recall of 72.93\% to locate the edit file, and
the precision of on average 86.97\% and the recall of 84.82\% to locate the edit lines.
We observe that the performance of \tool lies in identifying the edit pattern (e.g., the commit \texttt{4bf1c} in \texttt{Golang/Go} project (see an example at \cite{commit1}), which demonstrates the concrete example).
Further, the average runtime overhead to infer a file takes 1.6s.
We probe into the commits and summarize the reasons for false positives and false negatives as follows:

\noindent\textbf{Reason 1: Noisy Samples in the Training Dataset.}
As for inferring the location of subsequent edits,
we find that the quality of the dataset is of vital importance.
Despite that we have set a number of criteria to filter out some commits,
we still observe that noisy training samples might introduce negative effects.
One of the observations is that some programmers can submit some \textit{irrelevant} changes files (as well as the edits) in a single commit,
which makes \tool challenging to report some edit locations.
Further, we find that quite a number of edits are about code comments and documentation (e.g., the commit \texttt{3f442} in \texttt{golang/go} project \cite{commit2}),
which may not be well captured by \tool.

Nevertheless, cleaning the whole dataset regarding the edit relevance is a non-trivial work,
which is iterative and interactive between human observation/interpretation and automatic inference.
Thus, we leave the solution in our future work.

\noindent\textbf{Reason 2: Informativeness of Edit Inference.}
Further, we observe that some false negatives are caused by single-directed interaction.
For example, an addition of method call implies an addition of importing a relevant library declaring the method,
however, the implication does not hold in the other way (see example at \cite{commit3}).
When some interactions between the edits are not causal,
the inference becomes more challenging.

\begin{table}
\centering
\caption{The accuracy of propagating-file \& line location}
\label{tab:propagating-file}
\vspace{-5pt}
\small
\resizebox{\linewidth}{!}{%
\begin{tabular}{lccccc}
\hline
\multirow{2}{*}[-2ex]{\begin{tabular}[c]{@{}l@{}}\textbf{Programming }\\\textbf{Language}\end{tabular}} & \multicolumn{2}{c}{\textbf{File Location}} & \multicolumn{3}{c}{\textbf{Line Location}} \\ 
\cmidrule(l){2-3}\cmidrule(l){4-6}
 & \begin{tabular}[c]{@{}c@{}}\textbf{Precision}\\\textbf{(\%)}\end{tabular} & \begin{tabular}[c]{@{}c@{}}\textbf{Recall}\\\textbf{(\%)}\end{tabular} & \begin{tabular}[c]{@{}c@{}}\textbf{Accuracy}\\\textbf{(\%)}\end{tabular} & \begin{tabular}[c]{@{}c@{}}\textbf{Precision}\\\textbf{(\%)}\end{tabular} & \begin{tabular}[c]{@{}c@{}}\textbf{Recall}\\\textbf{(\%)}\end{tabular} \\ 
\hline
JavaScript                                                                                         & 81.52            & 71.21               & 94.89           & 86.62            & 83.88          \\
Python                                                                                             & 70.84            & 73.40               & 94.48           & 85.03            & 82.64          \\
Java                                                                                               & 85.28            & 75.67               & 95.37           & 87.99            & 85.99          \\
Go                                                                                                 & 80.10            & 72.12               & 95.79           & 88.99            & 87.32          \\
TypeScript                                                                                         & 79.84            & 72.25               & 95.23           & 86.21            & 84.25          \\
\hline
Average                                                                                            & 79.52            & 72.93               & 95.15           & 86.97            & 84.82          \\
\hline
\end{tabular}
}
\vspace{-5pt}
\end{table}


\subsubsection{RQ3 and RQ4 (Edit Generation \& Prior Edit Prediction)}\label{sec:rq3-res}
\autoref{tab:edit-generation} shows the overall performance of edit generation with 
Top-k candidates.
Further, \autoref{tab:prior-edit-on-locator} shows the relevance of prior edits in locating the subsequent edits and edit content generation.
We can see that
(1) \tool achieves good performance in generating the edit options, and
(2) the selective prior edits play a vital role in enhancing the performance.
An example can be referred to \cite{commit3},
where \tool is good at capturing the edit pattern (via syntactic dependency or semantic relevance).
The random prior edits can break the pattern,
which introduces additional edit chaos during the recommendation.
Further, we find that the mis-prediction of the edit options shares
similar reasons introduced in Section~\ref{sec:rq1-res}.

\begin{table}
\centering
\caption{The performance of edit generation}
\label{tab:edit-generation}
\vspace{-5pt}
\small
\begin{tabular}{lccccc}
\hline
\begin{tabular}[c]{@{}l@{}}\textbf{Programming }\\\textbf{Language}\end{tabular} & \textbf{Metric} & \textbf{Top-1} & \textbf{Top-3} & \textbf{Top-5} & \textbf{Top-10} \\
\hline
\multirow{2}{*}{Javascript} & \textbf{BLEU4} & 60.70 & 69.71 & 71.37 & 73.02 \\
 & \textbf{EMR(\%)} & 41.83 & 47.50 & 49.31 & 50.99 \\
\hline
\multirow{2}{*}{Python} & \textbf{BLEU4} & 57.59 & 65.65 & 67.47 & 69.11 \\
 & \textbf{EMR(\%)} & 33.48 & 38.52 & 40.41 & 42.09 \\
\hline
\multirow{2}{*}{Java} & \textbf{BLEU4} & 60.54 & 68.35 & 70.11 & 71.73 \\
 & \textbf{EMR(\%)} & 40.69 & 46.87 & 48.78 & 50.51 \\
\hline
\multirow{2}{*}{Go} & \textbf{BLEU4} & 65.37 & 71.96 & 73.47 & 74.98 \\
 & \textbf{EMR(\%)} & 48.94 & 55.09 & 57.18 & 59.16 \\
\hline
\multirow{2}{*}{Typescript} & \textbf{BLEU4} & 61.75 & 70.31 & 71.99 & 73.68 \\
 & \textbf{EMR(\%)} & 41.58 & 46.86 & 48.57 & 50.65 \\
\hline
\end{tabular}
\vspace{-5pt}
\end{table}

\begin{table}
\caption{Relevance of prior edits on edit location \& generation}
\label{tab:prior-edit-on-locator}
\vspace{-5pt}
\small
\centering
\begin{tabular}{lccccc}
\hline
\multirow{2}{*}[-1.5ex]{\begin{tabular}[c]{@{}l@{}}\textbf{Prior Edit} \\ \textbf{Relevance}\end{tabular}} & \multicolumn{3}{c}{\begin{tabular}[c]{@{}c@{}}\textbf{Edit-propagating} \\ \textbf{line locator}\end{tabular}} & \multicolumn{2}{c}{\begin{tabular}[c]{@{}c@{}}\textbf{Edit-content} \\ \textbf{generator}\end{tabular}} \\ 
\cmidrule(l){2-4}\cmidrule(l){5-6}
 & \begin{tabular}[c]{@{}c@{}}\textbf{Accuracy}\\\textbf{(\%)}\end{tabular} & \begin{tabular}[c]{@{}c@{}}\textbf{Precision}\\\textbf{(\%)}\end{tabular} & \begin{tabular}[c]{@{}c@{}}\textbf{Recall}\\\textbf{(\%)}\end{tabular} & \begin{tabular}[c]{@{}c@{}}\textbf{EMR}\\\textbf{(\%)}\end{tabular} & \textbf{BLEU4} \\ 

\hline
\begin{tabular}[c]{@{}l@{}}Selective \\ Prior Edits\end{tabular} & 94.89 & 86.62 & 83.88 & 41.83 & 60.70 \\
\begin{tabular}[c]{@{}l@{}}Random \\ Prior Edits\end{tabular} & 91.86 & 81.73 & 72.37 & 18.87 & 46.56 \\
\hline
\end{tabular}
\vspace{-5pt}
\end{table}

\subsubsection{RQ5 (Performance Boost)}\label{sec:rq5-res}
In \autoref{tab:performance-boost},
we compare \tool with three baselines, i.e., GRACE, CCT5, and CoditT5, in generating top-1 edit option.
As described in Section~\ref{sec:rq5-setup},
the fine-tuned baselines are fed with the hunk-level location (i.e., the lines included in a hunk) to predict
the edited code.
We can see that the performance gap between \tool and the baselines are large.
The reason lies in that the edit locator can largely help the edit generator to generate edits in a far more precise position.

Given that \tool is an extensible and integrable framework,
we replace our edit generation model with the fine-tuned baselines,
observing that the performance of both GRACE and CCT5 is boosted significantly.
Note that, CoditT5 can predict location as \tool,
we do not equip it with our locator.
In comparison to CoditT5,
we observe that our two-stage model has the advantage of utilizing more input length
given the limit of existing base models such as CodeT5.


\begin{table}
\centering
\caption{Performance Boost with \tool}
\vspace{-5pt}
\small
\label{tab:performance-boost}
\tabcolsep=0.1cm
\begin{tabular}{>{\hspace{0pt}}m{0.56\linewidth}>{\centering\hspace{0pt}}m{0.206\linewidth}>{\centering\arraybackslash\hspace{0pt}}m{0.127\linewidth}}
\hline
\textbf{Approach}                         & \textbf{EMR(\%)} & \textbf{BLEU4}  \\
\hline
CoEdPilot (Line Locator + Edit Generator) & \textbf{29.96} & \textbf{78.58}           \\
CoditT5                                   & 7.42         & 69.01           \\
GRACE                                     & 2.73         & 38.36           \\
CCT5                                      & 14.19        & 75.37           \\
GRACE + Line Locator                      & 18.61        & 71.61           \\
CCT5 + Line Locator                       & 15.45        & 78.27           \\
\hline
\end{tabular}
\vspace{-5pt}
\end{table} 
\section{User Study}\label{sec:user-study}

To further evaluate how the programmers can use \tool as a tool in practice,
we design a user study to evaluate its functionalities as described in Section~\ref{sec:tool}.

\noindent\textbf{Baseline.}
To evaluate whether the design of \tool can well support practical code edits,
we choose Copilot \cite{copilot} as a baseline for its wide popularity for generating code.
We omit the full manual editing mode in this study because
(1) Copilot is supported by powerful GPT-3.5 Turbo, which is shown to improve the programming productivity by 27\% - 57\% \cite{nguyen2022empirical}, and
(2) the limitation of budget and overhead.


\begin{table}
\centering
\caption{Runtime Estimation of \tool}
\label{tab:runtime}
\small
\vspace{-5pt}
\begin{tabular}{lccc}
\hline
\textbf{Step} & \begin{tabular}[c]{@{}c@{}}\textbf{File locator}\\~(s / file)\end{tabular} & \begin{tabular}[c]{@{}c@{}}\textbf{Line locator}\\~(s / file)\end{tabular} & \begin{tabular}[c]{@{}c@{}}\textbf{Edit-content} \\ \textbf{generator} \\ (s / location)\end{tabular} \\
\hline
Prepare Input & 0.0064 & 0.3976 & 0.0683 \\
Model Inference & 0.1008 & 0.0878 & 0.3972 \\
\hline
Total & 0.1072 & 0.4854 & 0.4655 \\
\hline
\end{tabular}
\vspace{-5pt}
\end{table}

\noindent\textbf{Participant.}
We recruit 18 participants from three universities in China and Singapore,
including both undergraduate and graduate students.
We conduct a pre-study (including a test) based on their programming experience.
Their demographic analysis is available at \cite{code-edit-pilot}.
We divide them into two equivalent groups based on their experience.
The experimental group uses \tool while the control group uses Copilot in the study.

\noindent\textbf{Code Edit Tasks.}
To ensure that the participants can focus on editing with a light-weighted overhead of comprehension,
we extract a simplified version from three real-world commits.
The tasks are selected as follows:
\begin{itemize}[leftmargin=*]
  \item \textbf{Bug Fix (Task 1)}:
    We show a bug as mistakenly used \texttt{range(arr)} for \texttt{range(len(arr))} in the project.
    We ask the participants to find and fix multiple such mistaken uses across the project.
  \item \textbf{Refactoring (Task 2)}:
    We ask the participants to extract three pieces of duplicated code into a new function.
  \item \textbf{Feature Enhancement (Task 3)}:
    We ask the participants to introduce a \textit{scale} capability to normalize the input vectors for existing class classifiers,
    which requires multiple edit propagation.
\end{itemize}

\noindent\textbf{Study Setup.}
We conducted a warm-up session with a tutorial for both \tool and Copilot,
followed by a practice task, to familiarize them with the tools.
For each task, we allocate each participant with 30 minutes to accomplish.
We prepare the test cases for each edit task for them to validate their edits.
The test cases are designed to guarantee that all the participants 
can confirm their accomplishing edits.
During the study, we ask the participants to run a video-recorder so that we can conduct the post-mortem analysis.
Finally, we measure their performance regarding
(1) whether they can successfully accomplish the tasks (i.e., all the test cases passed) and
(2) their efficiency in accomplishing the tasks.

\begin{table}
\centering
\caption{Overall performance of EG (Experimental Group) and CG (Control Group):The completion time is in seconds.}
\label{tab:us-performance}
\vspace{-5pt}
\tabcolsep=0.1cm
\small
\begin{tabular}{lccc|lccc}
\hline
\textbf{EG} & \textbf{Task1} & \textbf{Task2} & \textbf{Task3} & \textbf{CG} & \textbf{Task1} & \textbf{Task2} & \textbf{Task3} \\
\hline
\textbf{P1} & 221 & 515 & 1196 & \textbf{P10} & 339 & 696 & 1287 \\
\textbf{P2} & 897 & 389 & 279 & \textbf{P11} & 360 & 776 & 1563 \\
\textbf{P3} & 366 & 487 & 216 & \textbf{P12} & 480 & 483 & 545 \\
\textbf{P4} & 160 & 529 & 963 & \textbf{P13} & 522 & 724 & 1770 \\
\textbf{P5} & 230 & 301 & 756 & \textbf{P14} & 277 & 395 & 838 \\
\textbf{P6} & 364 & 473 & 617 & \textbf{P15} & 181 & 446 & 930 \\
\textbf{P7} & 329 & 688 & 588 & \textbf{P16} & 337 & 720 & 825 \\
\textbf{P8} & 840 & 780 & 1020 & \textbf{P17} & 151 & 666 & 1515 \\
\textbf{P9} & 290 & 638 & 1050 & \textbf{P18} & 266 & 722 & 1563 \\
\hline
\multicolumn{1}{c}{\textbf{Average}} & 410.78 & 533.33 & 742.78 & \multicolumn{1}{c}{\textbf{Average}} & 323.67 & 625.33 & 1070.33 \\
\hline
\end{tabular}
\vspace{-5pt}
\end{table}

\noindent\textbf{Results.}
\autoref{tab:us-performance} shows the participants' performance to accomplish the code-editing tasks, with the following observation:
\begin{itemize}[leftmargin=*]
  \item \textbf{Task 1}: EG underperforms CG in Task 1 on average completion time without statistical significance (the $p$-value in Wilcoxon Signed Rank test is 0.33 and the effect size is -0.08).
  \item \textbf{Task 2}: EG outperforms CG in Task 2 on average completion time without statistical significance (the $p$-value in Wilcoxon Signed Rank test is 0.07 > 0.05 and the effect size is 0.60).
  \item \textbf{Task 3}: In contrast, EG outperforms CG in Task 3 on the completion time with statistical significance (the $p$-value is 0.003 < 0.05 and the effect size is 0.96).
\end{itemize}

\noindent\textbf{Why CG outperforms EG in Task 1?}
In Task 1 (i.e., fixing a duplicated bug), we observe that \tool users (EG group) still suffer from the learning curve of the new tool for running the function of predicting edit location and edit content.
Further, some participants (P2 and P8) were still building their trust in our recommendations such as edit location and edit content,
despite that they are accurate.
As a result, they spend more time confirming our results.
We deem this a common problem for any new tool deployed on either user study or production line.
In contrast, given that the edit pattern in the task is simple, some Copilot user (e.g., P17) tries to search the expression across the project.
In the other words, they address their need of edit location with keyword-based search in Task 1.

\noindent\textbf{Why EG outperforms CG in Task 2 but without statistical significance?}
In Task 2 (i.e., refactoring by method extraction), the \tool users become more experienced in adapting our tool by switching between various functions such as location prediction, edit generation, edit option selection, etc.
The accurate edit location can largely mitigate the efforts in finding the cross-file code duplication for the new function.
Compared to the CG participants with manually summarized edit patterns,
the EG group gradually outperforms the CG group (the $p$-value 0.07 is closer to 0.05).

\noindent\textbf{How EG outperforms CG in Task 3?}
Task 3 (i.e., enhance the model training with \textit{scale} function) is the most difficult task,
where the editing pattern cannot be captured by keyword search.
For example, one edit to insert a \texttt{scale} parameter is associated with another edit to insert a follow-up decision logics with the \texttt{scale} variable.
In such a scenario, EG outperforms CG in general.

Nevertheless, we observe that the performance of the participants varies,
some accomplish the task in less than 5 minutes (e.g., P2) while some take a longer time (e.g., P1, P8, and P9).
We investigate their tool logs and videos,
finding that some participants modify the edit content with their own interpretation,
which leads to the buggy code.
Taking the buggy edit as the prior edits,
\tool can generate confusing edits afterwards.
Only by running the test cases to validate the results, 
the participants can realize they produce a bug during the editing.
Human mistakes in such an interaction-based tool are a long-standing problem,
we will address the issue in our future work.
Further, the \tool group accepts recommended 69.3\% edit options,
among which they modify 31.6\% generated edits.
For the space limit, more statistics of user behaviors in the study are available at \cite{code-edit-pilot}.



\section{Threat to validity}
Several aspects of the user study may impair its validity:

\noindent\textbf{Internal Validity:}
In this study, the experiment group may face a steeper learning curve,
while the control group is already acquainted with CoPilot.
This learning disparity could lead to observed differences in the test that are attributed to learning effects rather than the actual performance of the extension.

\noindent\textbf{External validity:}
The edit tasks are simplified versions of code derived from actual commits and equipped with comprehensive instructions.
This modification might deviate from the real editing scenario.
Moreover, as edit tasks exclusively focus on Python in this study,
such specificity choice could confound the interpretation of the plugin's effectiveness.

\noindent\textbf{Statistical Validity:}
Due to the limitation of time and resources, we recruited 18 participants in the study.
The relatively small size may not provide sufficient statistical power to detect genuine differences in the effectiveness of the extension.
Consequently, the generalizability and robustness of the study findings might be compromised.

\section{Related Work}

\noindent\textbf{Code Generation.}
Code generation is long standing software engineering task \cite{yang2021recent, yuan2023no, shin2021survey, cai2024fly, li2022soft},
which starts from sequence-based and tree-based approaches \cite{ling2016latent, sun2020treegen}, and gravitates towards pre-trained language models, such as BERT \cite{devlin2018bert}, GPT \cite{brown2020language}, T5 \cite{raffel2023exploring, han2021pre}, CodeBERT \cite{feng2020codebert}, GraphCodeBERT \cite{guo2020graphcodebert}, DietCodeBERT \cite{zhang2022diet}, CodeT5 \cite{wang2021codet5}, CodeT5+ \cite{wang2023codet5}, CodeT \cite{chen2022codet}, and Incoder \cite{fried2023incoder}.
Recently, StarCoder \cite{li2023starcoder} is trained with over 8 programming languages, Git commits, GitHub issues, and Jupyter notebooks.
It outperforms existing open Code LLMs on popular programming benchmarks and matches or surpasses closed models such as code-cushman-001 from OpenAI (the original Codex \cite{codex} model).
Meanwhile, our approach generates incremental edits, rather than new code.


\noindent\textbf{Code Edit Generation.}
Among the work to edit code \cite{CODIT, recoder, MODIT, overwatch, jiang2021cure, codereview, ZhangETAL22CoditT5, lin2015clone},
Codit \cite{CODIT} is the first to introduce tree-based neural networks for predicting the edits.
Following their tree model structure, Recoder \cite{recoder} introduces another abstract syntax tree (AST) reader along with the code reader to outperform the Codit model.
Further, CURE \cite{jiang2021cure} introduces the pre-training models for automatic program repair.
CoditT5 \cite{ZhangETAL22CoditT5} pre-train a CodeT5 \cite{wang2021codet5} base model with the input including natural language comments and edit code hunk and the output including an edit plan.
The current state-of-the-art transformer-based model is GRACE \cite{GRACE}, which trains a prompting large language model \cite{zhang2023prompting} with a designed prompt to include the associated code update.
Overwatch \cite{overwatch} symbolically analyzes edit sequence patterns by formulating them into rules based on prior program transformations.
Our solution, \tool, is complementary to the majority of the transformer-based code-edit generation model.
In addition to exploring the interactive nature of code edits,
we further learn to capture the relevant prior edits and subsequent edit location.

\section{Conclusion}
In this work, we introduce \tool, an end-to-end framework to interactively generate code edits by orchestrating a set of neural transformers as components, regarding prior edit analysis, subsequent edit analysis, and edit generation.
Our extensive experiments show that \tool is able to predict the edit location and generate edit options in an effective way.
Further, the framework is complementary to a set of state-of-the-art edit generators to boost their performance.
Our user study shows that \tool as a VS Code plugin is effective in assisting programmers in practice.
In the future, we will improve the quality of the training dataset for a more effective model and address the potential mistaken human feedback in tool design. 
\section{Data availability}
Our models and datasets are published on HuggingFace \cite{CoEdPilot}, both source code and VS Code extension are available on GitHub \cite{CoEdPilot_code, CoEdPilot_extension}. 

\begin{acks}
This research is supported in part by National Key Research and Development Program of China (Grant No.2023YFB4503802), the Bytedance Network Technology, the Minister of Education, Singapore (T2EP20120-0019, MOET32020-0004), the National Research Foundation, Singapore, and Cyber Security Agency of Singapore under its National Cybersecurity Research and Development Programme (Award No. NRF-NCR\_TAU\_\-2021-0002), National Research Foundation, Singapore, and the Cyber Security Agency under its National Cybersecurity R\&D Programme (NCRP25-P04-TAICeN), DSO National Laboratories under the AI Singapore Programme (AISG Award No: AISG2-GC-2023-008).
\end{acks}

\bibliographystyle{ACM-Reference-Format}
\bibliography{Reference}
\end{document}